\documentclass[letterpaper,onecolumn,preprintnumbers,amsmath,amssymb]{revtex4}
\usepackage[charter]{mathdesign}
\usepackage{siunitx}
\usepackage{graphicx}
\usepackage{wrapfig}
\usepackage{booktabs}
\usepackage{setspace}
\usepackage[unicode,colorlinks=true,citecolor=blue,urlcolor=blue,linkcolor=blue]{hyperref}
\setcitestyle{numbers,comma,sort&compress,super}
\usepackage{url}
\usepackage[normalem]{ulem}
\usepackage[nameinlink,capitalize]{cleveref}

\def\equationautorefname~#1\null{Equation~(#1)\null}
\usepackage{bbding}

\usepackage{ragged2e}
\newcommand{\ztitle}[1]{\title{\vspace*{3ex} \raggedright\fontfamily{mathfont}\Large\bf {#1}}}
\newcommand{\zauth}[1]{\author{\vspace{6ex}\hspace*{-50em} \RaggedRight\fontfamily{mathfont}\large\bf {#1}}}
\newcommand{\zaffi}[1]{\affiliation{\vspace{1ex} \raggedright\fontfamily{mathfont} {#1}}}
\begin{document}
\ztitle{Ultimate speed of the supercurrent and its pairing mechanism}
\zauth{N. Zen}
\zaffi{Device Technology Research Institute, National Institute of Advanced Industrial Science and Technology,\\Tsukuba Central 2-10, Ibaraki 305-8568, Japan\vspace{1ex}\\
  {\it E-mail address:} {\tt\small n.zen@aist.go.jp}\\
  {\it ORCID ID:} \href{https://orcid.org/0000-0003-4897-478X}{\tt\small 0000-0003-4897-478X}
\vspace{1ex}}
\date{\today}
\maketitle
\onecolumngrid
\vspace{-1.5ex}
\hrule
\vspace{1ex}
\noindent
{\large\bf Abstract}\vspace{0.5ex}\\
Recently, the room-temperature superconductor (RTSC) was discovered as a two-dimensional (2D) square lattice made of a metal wherein positive charges, i.e. holes, were heavily concentrated. The experimental result for the critical magnetic field $H_{c}$ was fully consistent with the view on the RTSC that its lattice unit---a metal island---is filled with the Slater's atoms. Each Slater's atom has the expanded diameter of 14.5 nm in order to have perfect diamagnetism with the magnitude corresponding to a single flux quantum $\phi_{0}$. Its expanded orbit is associated with the fine structure constant $\alpha \approx \frac{1}{137}$. In this paper, another important critical value---the critical current $I_{c}$---is reported. It was found that the supercurrent has achieved the ultimate speed of matter, i.e., the speed of light, $c$. Beginning with a warm-up exercise for the Bohr's atom, how the Slater's atom is formed and why the $c$ appears are shown. These considerations lead to a simple view on the pairing mechanism of superconductivity, which also gives an ample indication of the most mysterious physical number $\alpha$. Finally, it is shown that the proposed pairing mechanism in terms of London's canonical momentum naturally generates the perfect diamagnetic $\phi_{0}$ of the Slater's atom, and the superconducting energy gap $\mathnormal{\Delta}$ is predicted.
{\if0 At the end of this paper, an early result of a new solid state, which emerged after the excessive injection of charge carriers, is shown. The newly emergent semiconducting--superconducting junction shows diode effects and phase slip phenomena, justifying the concept of the giant atom that is to have an extraordinarily long superconducting coherence length ($\xi$) being equivalent to the macroscopic lattice constant of this sample, $a$ of 20 $\si{\micro m}$.\fi}\vspace{1ex}
\hrule
\vspace{2.5ex}
\noindent
{\it Keywords:}\\{\tt\small Room-temperature superconductor, Critical current $I_{c}$, Penetration depth $\lambda_{\perp}$, Slater's atom,\\Fine structure constant $\alpha$, Faraday's electric field in 2D, Superconducting energy gap $\mathnormal{\Delta}$}
\clearpage
\section{Introduction}
\label{sec:intr}
In the previous study~\cite{RTSC}, resistive and magnetic properties of a periodically perforated metal sheet designed by phonon engineering~\cite{Zen2014} were investigated. During temperature cycles, the metallic two-dimensional (2D) square lattice exhibited both resistance drop to zero and magnetization drop, i.e., the Meissner effect. Thus it was concluded that the 2D square lattice underwent a superconducting transition during the temperature cycle. The most astonishing result of the experimental study was that the 2D square lattice retained the superconductivity at higher temperatures in a warming process than that in a cooling one. Despite the general belief that lower temperatures are always beneficial to superconductivity, the universal Ginzburg-Landau (G-L) theory by contrast teaches us that a warming process is always beneficial to superconductivity whenever it shows thermal hysteresis as can be learnt from ref.~\cite{HMarxiv2021Jan}.
\vspace{1ex}

\begin{figure}[b!]
\centering
\includegraphics[width=\textwidth]{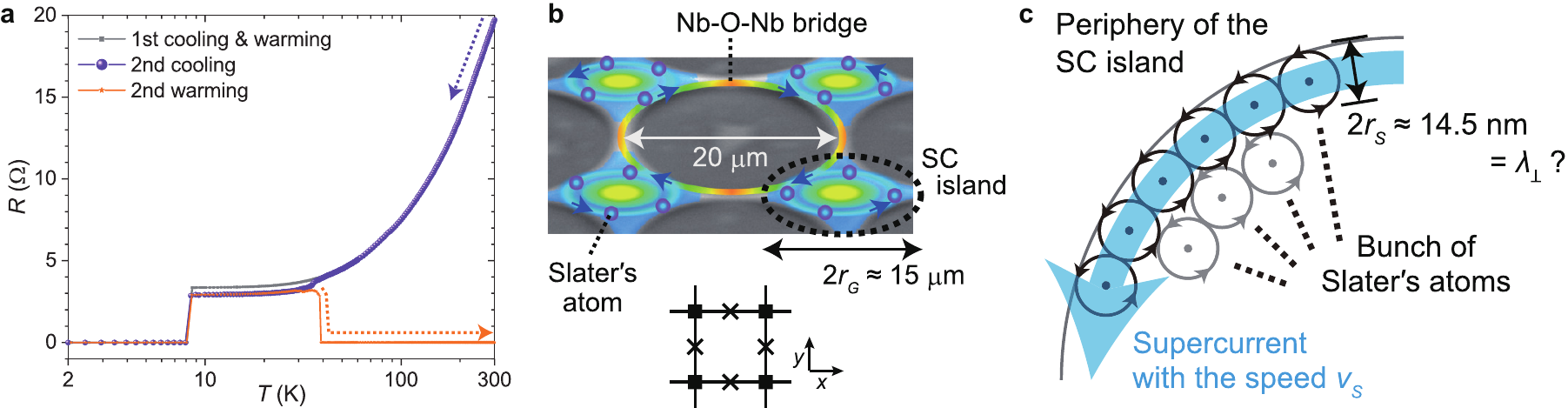}
\caption{\textbf{Room-temperature superconductor (RTSC) and its microscopic perspective.} \textbf{a}, \emph{R--T} characteristics of an RTSC sample used for this study. \textbf{b}, A false-colour SEM of a part of the sample, showing four Nb superconducting (SC) islands with the diameter $2r_{G}\approx 15~\si{\micro m}$. Each SC island is coupled to its four nearest neighbors by an Nb--O--Nb bridge. Bottom panel, a possible diagram of the RTSC sample. The boxes represent SC islands, and the symbols ``$\times$'' denote the Josephson junctions (JJs) joining them. \textbf{c}, Schematic illustration of the periphery of the SC island, wherein Slater's atoms with the radius $r_{S}$ are arranged in such a way as to be associated with the critical state of the RTSC. Cyan arrow, a supercurrent circulating around the periphery of the SC island. If this illustration is true, the penetration depth $\lambda_{\perp}$ must correspond to $2r_{S}\approx 14.5~\si{nm}$.
\label{fig1}}
\end{figure}

\autoref{fig1}a shows temperature dependence of resistance of the 2D square lattice used for this study. In the second warming process (orange), the resistance dropped to zero at 39 K, and the zero-resistance state was retained with the temperature warmed up to 300 K. Thus it has been revealed again that the warming process is indeed beneficial to superconductivity, and in this way the room-temperature superconductor (RTSC) has arrived. Besides the phenomenological explanation given by the G-L theory, the dynamics of the superconducting transition as well was discussed in the previous paper~\cite{RTSC} based on the Hirsch's theory of hole superconductivity~\cite{HirschBook,HirschWeb,HirschJAP2021}. The theory teaches us the absolute benefit of lattice expansion that lowers the orbital kinetic energy of an atom at each lattice point. When the self-standing 2D square lattice is expanded in the warming process, the expansion lowers the orbital kinetic energy of each atom,
\begin{equation}
E_{kin}=\frac{\hbar^{2}}{2mr^{2}},\label{eq:ekin}
\end{equation}
where $\hbar$ is the reduced Planck constant, $m$ is the mass of the orbital charge and $r$ is the orbital radius. When the kinetic energy is thus lowered by the orbital expansion, all the atoms achieve kinetic equilibrium, then, they go superconducting. This is consistent with the London's consideration about superconducting transition: ``\emph{the most stable state of any system is not a state of static equilibrium in the configuration of lowest potential energy. It is rather a kind of kinetic equilibrium ... .}''~\cite{LondonBook} Any superconductor is in kinetic equilibrium, and, conversely, kinetic equilibrium triggers superconductivity.
\vspace{1ex}

According to London and Slater~\cite{London1937,Slater1937}, the expanded atom is the element of any superconductor. It cannot be separated into simpler units any further. It has perfect diamagnetism with the magnitude corresponding to a single flux quantum $\phi_{0}$. For this, Slater calculated the orbital radius of the expanded atom as
\begin{equation}
r_{S}\approx 137\times a_{0}~~(\approx \mathrm{7.25~nm}),\label{eq:rs}
\end{equation}
where $a_{0}$ is the Bohr radius (0.529 \AA) and the number 137 comes from the fine structure constant ($\alpha\equiv \frac{e^{2}}{\hbar c}\approx \frac{1}{137}$). In other words, the expanded atom is approximately 137 times larger than the Bohr's atom. It is called the Slater's atom.
\vspace{3ex}

For the sake of the diamagnetic $\phi_{0}$, Slater gave his atom the expanded radius $r_{S}\approx 137a_{0}$. Without $\phi_{0}$, however, he could have provided neither $r_{S}$ nor the number 137. This is obvious from his question: ``\emph{why we could not let the size of our cells increase, from 137 atom diameters, to any desired size.}''~\cite{Slater1937} First, this paper will answer his question by considering the experimental result for the critical current $I_{c}$ without assuming $\phi_{0}$. Despite the fact that the physical meaning of the number 137 is even now shrouded in mystery, it turns out that the answer for his question is very simple.

\section{Experimental critical current and its speed}
\label{sec:ic}
The critical magnetic field $H_{c}$ for the RTSC at 300 K was experimentally proven to be a little bit larger than 12 T~\cite{RTSC}, and the value turned out to be consistent with a simple equation,
\begin{equation}
\mu_{0}H_{c}=\frac{\phi_{0}}{\pi\times r_{S}^{2}}~~(\approx \mathrm{12.5~T}),\label{eq:hc}
\end{equation}
which brought a microscopic perspective on the RTSC as illustrated in \cref{fig1}c. At the critical state the RTSC is filled with a bunch of Slater's atoms overlapping each other. This is what \cref{eq:hc} is indicating and why the RTSC has such a large $H_{c}$. In the inner domain of the illustrated superconducting (SC) island, the superposition of orbital charges nullifies themselves, leaving orbits only within a thin layer adjacent to the periphery. In other words, the width of a supercurrent path is $2r_{S}\approx 14.5~\si{nm}$ either it is a screening current under applied field or the critical current in the absence of applied field. The width of the former current is in particular called the penetration depth $\lambda_{\perp}$.
\vspace{1ex}

\begin{wrapfigure}{r}{0.45\textwidth}
\centering
\includegraphics[width=0.45\textwidth]{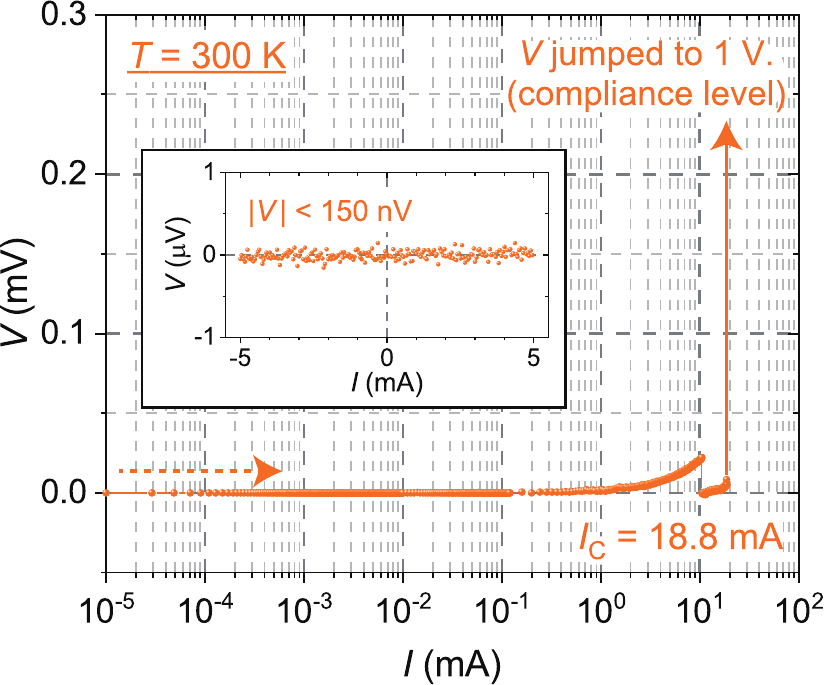}
\caption{\textbf{Critical current measured at 300 K.} By applying current, the output voltage was measured at 300 K in the absence of applied magnetic field. Inset, the result for both polarities of the applied current with its magnitude lower than $I_{c}$.
\label{fig2}}
\end{wrapfigure}

The critical current $I_{c}$ of the RTSC was investigated at 300 K in the absence of applied field by applying current and measuring the output voltage. As shown in the inset of \cref{fig2}, the output voltage is in the lowest limit of accuracy of the measurement system for both polarities of applied current with its magnitude lower than 5 mA. Thus the sample is indeed in the superconducting state at 300 K. Subsequently, DC current was applied to the RTSC and was increased from $+0~\si{mA}$ as shown in the main panel. The output voltage starts to increase slightly for DC currents exceeding $+1~\si{mA}$, which might be due to a thermal voltage of aluminium wires conducting electricity to the sample. Finally, at $+18.8~\si{mA}$, the voltage jumped to $1~\si{V}$ which was the set compliance level of the voltmeter. Thus, the $I_{c}$ of the RTSC at 300 K is $18.8~\si{mA}$.
\vspace{1ex}

For a 2D Josephson-junction (JJ) array in a square-lattice configuration, the penetration depth $\lambda_{\perp}$ of the system is given by~\cite{Phil1993,TinkhamBook}$^{[Section~6.6]}$
\begin{equation}
\lambda_{\perp}=\frac{\phi_{0}}{2\pi\mu_{0}\times I_{c}}.\label{eq:lambda}
\end{equation}
By substituting the $I_{c}$ of the RTSC (18.8 mA), the $\lambda_{\perp}$ is calculated as 14.0 nm which is roughly consistent with the $2r_{S}\approx 14.5~\si{nm}$. Thus, the aforementioned assumption---the width of a supercurrent path is $2r_{S}$---and its associated perspective on the RTSC (\cref{fig1}c) are validated. Note that both \cref{eq:hc,eq:lambda} are valid only in 2D cases. This indicates that the RTSC in this study is indeed a `2D superconductor', which must reflect a key role of 2D phonon engineering~\cite{Zen2014} that nullifies 3D components of electron momentum as explained before~\cite{RTSC}. Another indication of \cref{eq:lambda}, i.e. the `JJ array', may be understood by taking a look at \cref{fig1}b wherein four SC islands are depicted. The previous study using energy dispersive x-ray (EDX) imaging~\cite{RTSC} revealed that oxygen atoms highly invaded the bridge parts joining SC islands. Hence it must be plausible to assume that the Nb--O--Nb bridge is functioning as a Josephson weak link and that the RTSC is forming a 2D JJ square array. At least its $I_{c}$ satisfies the \cref{eq:lambda} which is valid for a 2D JJ square array. Properties of the RTSC as a JJ array are however not the main topic of this study. Those will be adequately dealt with in the next paper.
\newpage

Given that the 2D supercurrent $I_{S}$ has the width $2r_{S}$ and an infinitesimal thickness $\delta z$ and is carried by a pair of the elementary charge $q\approx 1.602\times 10^{-19}~\si{\coulomb}$ with a speed $v_{S}$, the $I_{S}$ is expressed as
\begin{equation}
I_{S}\equiv 2q\times n\times v_{S}\times S=2q\times \frac{2}{\pi r_{S}^{2}\times \delta z}\times v_{S}\times 2r_{S} \delta z=\frac{8q}{\pi r_{S}}\times v_{S}\approx \frac{8q}{\pi\times 137a_{0}}\times v_{S},\label{eq:ic}
\end{equation}
where $n$ and $S$ is carrier density and cross section of the current path, respectively. When the RTSC is going to the critical state, the $I_{S}$ approaches the critical value $I_{c}$ of 18.8 mA. Then the $v_{S}$ is approximately $3\times 10^{8}~\si{m/s}\approx c$, the speed of light. That is, at the critical state, the charge pair is orbiting around the Slater's atom at the ultimate speed of matter. Because nothing can leave light behind, the Slater's atom cannot expand its orbit any longer. This is the answer to the Slater's question---``\emph{$c$ restricts your radius to $137a_{0}$.}'' The relation between the orbital radius and the orbital velocity, together with why and how the orbit expands, are explained in the following sections using only elementary physics. Regarding \cref{eq:ic}, some smart readers may question whether the former part ``$2q\times n$'' should be ``$q\times n$'' or ``$2q\times n/2$''. Yet \cref{eq:ic} is correct. The reason for this can be found in the last part of this paper. Let me begin with the formation of the Bohr's atom.
{}

\section{Theory}
\label{sec:theo}
\subsection{Bohr's atom}
\label{sec:bohr}
When there are two particles with opposite charges in space, they attract each other through electromagnetic force $F_{E}$ under Faraday's principle of locality (\cref{fig3}a). And Newton's first law demands that they continue attracting each other; in other words, each particle has to continue falling into the other for ever. Unless they annihilate each other, the only possible solution is to orbit each other as shown on the bottom panel of \cref{fig3}a.{\if0 This is the very moment the physical number $2\pi$ is generated.\fi} In this manner, both particles can fall for ever. Note that this view tacitly agrees with the cosmological principle---\emph{no place in the universe is privileged}. This in turn indicates that the Bohr model does not agree with the cosmological principle because it sets a positive charge at the privileged origin of the universe. If his model is correctly redrawn as shown, the Larmor electromagnetic radiation loss due to the accelerated motion of a charge is no longer necessary to be taken into account. Each Larmor radiation is properly absorbed by its orbit, and the total energy of the system within the two orbits is properly conserved.
\vspace{1ex}

When the distance between two charges at points at infinity approaches an equilibrium distance $r_{0}$, the work done in this process is
\begin{equation}
W=\int_{\infty}^{r_{0}}F_{E}(r)dr=\int_{\infty}^{r_{0}}\frac{q^{2}}{4\pi \epsilon_{0}}\frac{1}{r^{2}}dr=-\frac{q^{2}}{4\pi \epsilon_{0}}\frac{1}{r_{0}},\label{eq:w}
\end{equation}
where $\epsilon_{0}\approx 8.85\times 10^{-12}~\si{F/m}$ is the electric constant. The negative $W$ indicates that the two charges indeed prefer not to stay at points at infinity. This in turn indicates that the system gains the amount of energy corresponding to $W$. By considering that the gained energy is converted to the kinetic energy of the two orbits each of which is given by \cref{eq:ekin},
\begin{equation}
\frac{q^{2}}{4\pi \epsilon_{0}}\frac{1}{r_{0}}=2\times E_{kin}=2\times \frac{\hbar^{2}}{2mr_{0}^{2}},\label{eq:w2kin}
\end{equation}
which gives the solution,
\begin{equation}
r_{0}=\frac{4\pi \epsilon_{0} \hbar^{2}}{mq^{2}}\equiv a_{0}\approx 0.529~\si{\angstrom}.\label{eq:a0}
\end{equation}
Thus the Bohr radius $a_{0}$ is properly obtained.
\vspace{1ex}

However, the proposed view (bottom panel of \cref{fig3}a) is not yet accurate. Each orbiting particle shall fail to receive the other's Larmor radiation. This difficulty was resolved thanks to de Broglie. Each orbit should be considered as a wave instead of a particle's track. By equating \cref{eq:ekin} as
\begin{equation}
E_{kin}=\frac{\hbar^{2}}{2mr^{2}}=\frac{1}{2}mv^{2},\label{eq:ekinparticle}
\end{equation}
where $v$ is the orbital velocity, this yields the orbital kinetic momentum,
\begin{equation}
p\equiv mv=\frac{\hbar}{r}=\frac{h}{2\pi r}=\frac{(the~Planck~constant)}{(the~orbit's~circumference)}.\label{eq:p}
\end{equation}
In turn, \cref{eq:ekin} is rewritten as
\begin{equation}
E_{kin}=\frac{\hbar^{2}}{2mr^{2}}=\frac{1}{2}mv^{2}=\frac{1}{2}pv=\frac{1}{2}\times \frac{h}{2\pi r}\times v.\label{eq:ekinwave}
\end{equation}
Both the final forms of \cref{eq:p,eq:ekinwave} do not include ``$m$'' explicitly. Hence we can now deal with an orbit properly in wave view. For exercise, \cref{eq:w2kin} with the already obtained solution ($r=a_{0}$) can be rewritten as
\begin{equation}
\frac{q^{2}}{4\pi \epsilon_{0}}\frac{1}{a_{0}}=2\times E_{kin}=\frac{h}{2\pi a_{0}}\times v,\label{eq:w2kin2}
\end{equation}
which yields the orbital velocity,
\begin{equation}
v=\frac{q^{2}}{2\epsilon_{0} h}\equiv v_{B}\approx 2.19\times 10^{6}~\si{m/s}.\label{eq:vb}
\end{equation}
Thus the Bohr velocity $v_{B}$ is properly obtained. It is also noteworthy that the $v_{B}$ can be expressed as
\begin{equation}
v_{B}\equiv \frac{q^{2}}{2\epsilon_{0} h}=\frac{q^{2}}{2\epsilon_{0} h c}\times c\equiv \alpha\times c\approx \frac{1}{137}c.\label{eq:vbc}
\end{equation}
For the fourth formula, the definition of $\alpha$ in SI units is used. \autoref{eq:vbc} indicates that the Bohr orbital velocity is approximately 137 times slower than the speed of light. This is a well-known fact but has never been discussed before in terms of superconductivity.

\subsection{Orbital escape velocity}
\label{sec:escape}
So far, it is not clearly stated what these two particles are. That is, all the above equations are valid no matter what they are. They can be an electron, a proton, a positron, a hole---a `hole' in an electronic band, etc. A nucleus---the pack of a proton and a neutron---is also possible since ``$m$'' is not included anywhere in \cref{eq:p,eq:ekinwave}.{\if0 Only its orbital radius and velocity are necessary to understand the physics of the system.\fi} Although there is the history that the Bohr model originated with Rutherford's experiments, the model is in fact not restricted to the combination of a nucleus and an electron. It is applicable to any combination of the above, only if the two particles have opposite charges{\if0 The inevitable condition for the model is that they have to have opposite charges\fi}. Beyond probability, Newton's first law pronounces their fate, decreeing that they each orbit the other. They each do not need any information other than the charge of the other. They each orbit around whatever the other is, no matter what fine structure the other has.
\vspace{1ex}

\begin{figure}[b!]
\centering
\includegraphics[width=0.8\textwidth]{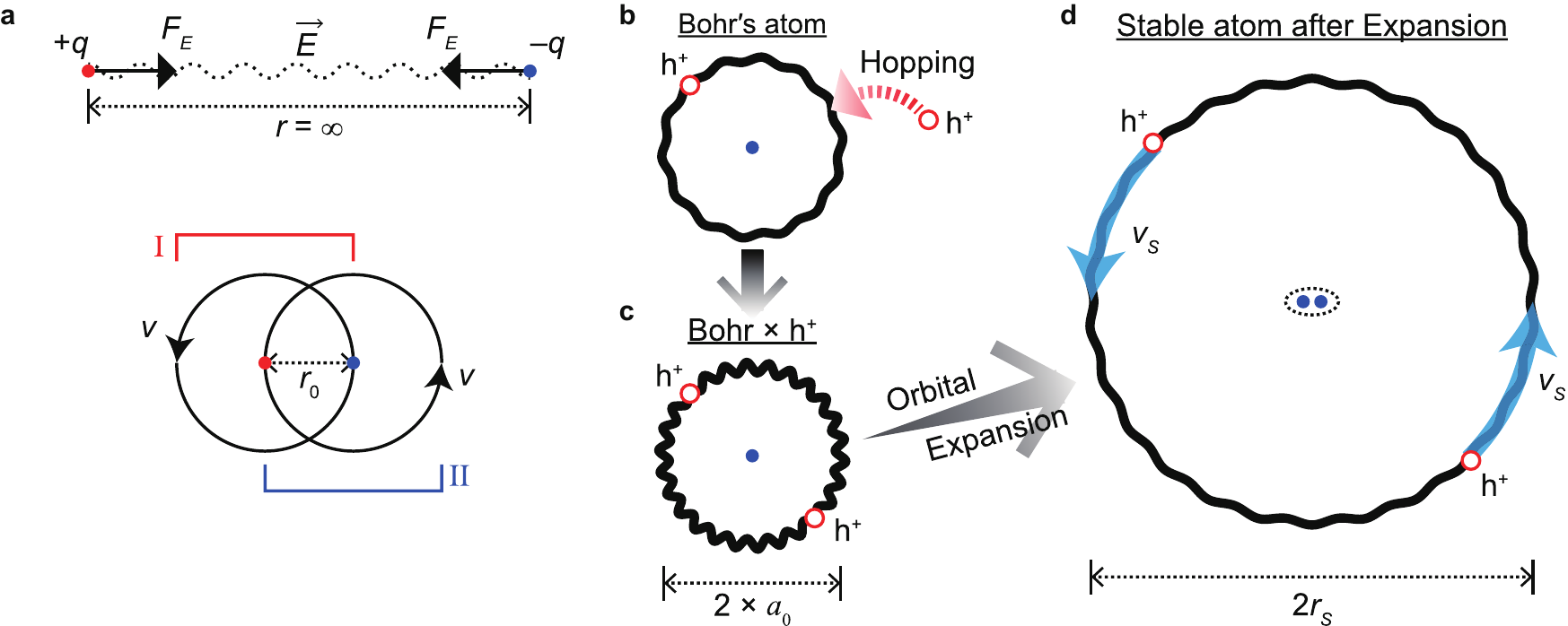}
\caption{\textbf{Bohr to Slater's atom.} \textbf{a}, Two particles with opposite charges approaching each other, forced by $F_{E}$ in the electric field $\vec{E}$ under Faraday's principle of locality. ``$q$'' denotes the elementary charge, approximately $1.602\times 10^{-19}~\si{\coulomb}$. Bottom panel, its consequence by taking into account the cosmological principle---\emph{no place in the universe is privileged}. As there is no privileged origin of coordinates, two orbits have to be illustrated. At equilibrium, the orbital radius $r_{0}$ and velocity $v$ correspond to those of the Bohr's atom. \textbf{b}, Stable Bohr's atom whose conduction band is almost full of electrons and hence its electrical conductivity is by a hole ($\rm{h}^{+}$). \textbf{c}, Destabilized Bohr's atom by the hopping of an $\rm{h}^{+}$ into the same orbit. \textbf{d}, Stable atom after the orbital expansion. The orbital radius and velocity is denoted by $r_{S}$ and $v_{S}$, respectively. Blue dots, negative charges that stabilize the atom.
\label{fig3}}
\vspace{-4ex}
\end{figure}

Yet the proposed view considering two atoms at the same time (bottom panel of \cref{fig3}a) is inconvenient to proceed with. The atoms ``I'' and ``II'' are equivalent to each other in the name of the cosmological principle as mentioned before. From now on, only the atom ``II'' wherein a positive charge orbits a negative charge is considered for the sake of superconductivity. This study is following Kikoin, Born, Feynman, Chapnik {\it et al.}~\cite{Kikoin1932,Born1948,Feynman1957,Chap1979} and Hirsch~\cite{Hirsch1989} who have concluded that electrical conductivity of any superconductor is not by electrons but by holes. Therefore the atom ``II'', wherein a positive charge orbits and hence conducts electricity, will lead this study to the right conclusion. Because a hole ($\rm{h}^{+}$) is a `hole' in an almost full electronic band, the atom ``II'' is redrawn as \cref{fig3}b wherein an almost full electronic band and an $\rm{h}^{+}$ in the band is represented by the bold wavy track and the empty dot in the track, respectively. De Broglie's wave view is still preserved. That is, we continue to use \cref{eq:p,eq:ekinwave} to describe the physics of the positive orbit.
\vspace{1ex}

The negative charge on the other hand is now settled at the core of the atom, i.e., at the privileged origin of the universe. As stated above, it does not matter what the negative charge is. Therefore we forget about what it is for now. It will be identified later in this paper. Additionally, Larmor radiation remains not to be taken into account. If anyone cares about radiation loss, let the core orbit around the $\rm{h}^{+}$, then, the radiation term will disappear.
\newpage

Now the system has lost one of two orbits because of the settlement at the privileged origin (\cref{fig3}b). The system thus reduced its total energy by $1\times E_{kin}$. However, the settled negative point charge at the core creates a static electric field instead, which is given by Gauss's law as
\begin{equation}
-\frac{q}{\epsilon_{0}}=\int_{S}\vec{E}(r)\cdot\vec{n}(r)dS,\label{eq:gauss}
\end{equation}
where $\vec{E}(r)$ is an electric field at the distance $r$ from the core and $\vec{n}(r)$ is the normal unit vector. By solving this surface integral for a 3D shell having the surface area $4\pi r^{2}$, one obtains the magnitude of $\vec{E}(r)$ as
\begin{equation}
E(r)=-\frac{q}{4\pi \epsilon_{0}} \frac{1}{r^{2}}.\label{eq:e3d}
\end{equation}
Using the definition for the electrostatic potential $\phi(r)$, namely $\vec{E}(r)\equiv -\rm{grad}~\it{\phi(r)}$, the 3D $\vec{E}(r)$ corresponding to \cref{eq:e3d} yields
\begin{equation}
\phi(r)=-\frac{q}{4\pi \epsilon_{0}} \frac{1}{r}.\label{eq:v3d}
\end{equation}
This in turn gives a potential energy to the positive orbit which is closing the shell using its own orbital radius $a_{0}$,
\begin{equation}
E_{pot}=q\times \phi(a_{0})=-\frac{q^{2}}{4\pi \epsilon_{0}} \frac{1}{a_{0}}.\label{eq:epot}
\end{equation}
Although $E_{pot}$ has the same formula as the work $W$ given by \cref{eq:w}, the physical origin is completely different. $W$ arose from 1D motions of two particles. On the other hand, $E_{pot}$ appears \emph{only after} the higher-dimension (2D or 3D) shell is formed. Most introductory textbooks explain that $W$ gives $E_{pot}$ of an atom. But Newton does not teach in such a slipshod manner. Two particles have to form orbits first. Then if either of two particles is set at the origin, the other gains $E_{pot}$, which originates from the Faraday's electric field that is effective \emph{only within} the inner shell of the atom. Also, such a direct conversion from $W$ to $E_{pot}$ cannot generate any single orbital $E_{kin}$ anywhere in the atom. That is, such a direct conversion is violating the law of conservation of energy; nonetheless, it is generally accepted---we have to choose which side to take. The forthright view on an atom is that (i) only its orbit is important, no matter what is at the core, (ii) its kinetic momentum $p$ is determined only by the orbit's circumference in accordance with \cref{eq:p}, no matter how ``heavy'' the orbit is, and (iii) its kinetic energy $E_{kin}$ is given by $\frac{1}{2}pv$ where $v$ is the orbit's velocity in accordance with \cref{eq:ekinwave}.
\vspace{2ex}

In celestial mechanics, once an object has achieved an escape velocity, it cannot stay in a closed orbit of any radius any longer, moving away from the gravitational influence of a primary body, never to come back. The escape velocity is easily calculated by considering an object's kinetic energy that exceeds its gravitational potential energy. When this procedure is applied to the $\rm{h}^{+}$-orbit with the radius $r=a_{0}$ and hence whose kinetic momentum $p$ is always fixed to the value $\frac{h}{2\pi a_{0}}$, the escape velocity $v_{es}$ of the $\rm{h}^{+}$-orbit is given by the condition,
\begin{equation}
E_{kin}=\frac{1}{2}\times \frac{h}{2\pi a_{0}}\times v_{es}\geq |E_{pot}|=\frac{q^{2}}{4\pi \epsilon_{0}} \frac{1}{a_{0}},\label{eq:ekinepot}
\end{equation}
which gives the solution,
\begin{equation}
v_{es}\geq \frac{q^{2}}{\epsilon_{0} h}=2v_{B}.\label{eq:vs}
\end{equation}
For the last term, the definition of the Bohr velocity $v_{B}$ in \cref{eq:vb} is used. This consequence indicates that when the orbital velocity gets two times faster, the orbit cannot stay at $r=a_{0}$ any longer and moves away, never to come back.
\vspace{2ex}

Which physical situation meets the condition \cref{eq:ekinepot}? Let's assume, for instance, the situation where two $\rm{h}^{+}$-orbits with $r=a_{0}$, hence $p=\frac{h}{2\pi a_{0}}$, and the orbital velocity $v_{B}$ are superpositioned. When two identical waves are superpositioned, the emergent new wave will have two times larger kinetic momentum and in turn two times larger kinetic energy. In the case of $\rm{h}^{+}$-orbits, the superposition will yield the new kinetic energy,
\begin{equation}
E_{kin}'=\frac{1}{2}\times 2p\times v_{B}=\frac{1}{2}\times p\times 2v_{B}.\label{eq:ekin'}
\end{equation}
As the last formula is equivalent to the left-hand side of \cref{eq:ekinepot}, the superpositioned two $\rm{h}^{+}$-orbits will move themself away from the electrodynamic influence of the core point charge. In other words, the superpositioned orbit starts to expand. However, it does not expand to infinity. There is a limit ``$137a_{0}$'' as mentioned before. It is the secret of $\alpha\approx \frac{1}{137}$, which is explained in the next section.
\vspace{1ex}

By the way, the assumed superpositioned situation can be illustrated as \cref{fig3}c. The doubly wave-numbered track represents the superpositioned $\rm{h}^{+}$-orbit having the double kinetic momentum $2p=2\times \frac{h}{2\pi a_{0}}$. Nevertheless its radius is still kept at $r=a_{0}$. That is, the orbit is very unstable and is about to expand as calculated above. In particle view, this situation corresponds to a Bohr's atom which contains an $\rm{h}^{+}$-pair in its single orbit. Hirsch has developed his theory of hole superconductivity using a model, which he calls the dynamic Hubbard model~\cite{HirschDHM2013}. In contrast to conventional Hubbard models, the dynamic Hubbard model \emph{does} take into account the fact that the wavefunction of an atomic orbit expands when a second charge occupies the orbit. This fact is significant especially for a material whose electrical conductivity is by holes. Because the material's electronic band is almost full of electrons, there is almost no available energy level in the band for a second hole from the outside of the band to hop into{\if0 or an already existing hole to escape into\fi}. Nevertheless, if the second hole overcomes this strongly repulsive situation by some means and hence hops into the band, the band gets catastrophically unstable and is consequently going to expand. This consideration is consistent with the truth revealed by Kikoin, Born, Feynman, Chapnik {\it et al.}~\cite{Kikoin1932,Born1948,Feynman1957,Chap1979} that holes conduct supercurrent. This holds true also for copper-oxide superconductors according to M\"{u}ller. In 1987 he remarked ``\emph{This new class of materials found at the IBM Zurich Research Laboratory are hole rather than electron superconductors.}''~\cite{Muller1987} Twenty-seven years later in 2014, he still defined ``\emph{Cooper pairs}'' as ``\emph{two holes with antiparallel spins}.''~\cite{Muller2014} Maybe he will do so even now. Then, by what means does the second hole overcome such a strong repulsive situation and occupy the same orbit with its partner? It could be chemical hole doping, physical hole currents repeatedly applied as was done for this study~\cite{RTSC} or oxygen transformation associated with the Jahn-Teller theorem~\cite{Muller2008}. By whatever means, once an extra $\rm{h}^{+}$ hops into an $\rm{h}^{+}$-orbit, the $[\rm{h}^{+}\!\times\rm{h}^{+}]$-orbit gets unstable, expands and goes superconducting. This last remark is also consistent with the Matthias's conviction---``\emph{Instabilities associated with high transition temperature superconductors}''~\cite{Matt1973}---of course.

\subsection{Kinetic equilibrium}
\label{sec:slater}
Kinetic energy of the unstable $[\rm{h}^{+}\!\times\rm{h}^{+}]$-atom (\cref{fig3}c) and that of the stable one after the expansion (\cref{fig3}d) is
\begin{subequations}\label{all}
\vspace{-1ex}
\begin{align}
&E_{kin}'=\frac{1}{2}\times 2p\times v_{B}=\frac{h}{2\pi a_{0}}\times v_{B}\approx \frac{h}{2\pi a_{0}}\times \frac{1}{137}c,\label{eq:ekin'2}\\
&E_{kin}^{(S)}=\frac{1}{2}\times 2p_{S}\times v_{S}=\frac{h}{2\pi r_{S}}\times v_{S},\label{eq:ekinS}
\end{align}
\end{subequations}
respectively. For each equation, double kinetic momentum is used since each orbit is a doubly superpositioned one; in terms of particle, each equation reflects the fact that each orbit includes two $\rm{h}^{+}$'s. The definition of kinetic momentum can be found in \cref{eq:p}. For the last term of \cref{eq:ekin'2}, the definition of $v_{B}$ in \cref{eq:vbc} is used. At present we still do not know what the stable atom is, namely, its orbital radius $r_{S}$ and orbital velocity $v_{S}$.
\vspace{1ex}

Assuming \cref{eq:ekinS} is the equation for a superconducting atom, we can equate those equations as $E_{kin}'=E_{kin}^{(S)}$ in the name of London and Hirsch's \emph{kinetic} equilibrium as \emph{the} principle of superconducting transition~\cite{HirschBook,HirschWeb,HirschJAP2021,LondonBook}. By taking into account the experimental result $v_{S}\approx c$ given by \cref{eq:ic}, the kinetic equilibrium yields
\begin{equation}
\frac{h}{2\pi a_{0}}\times \frac{1}{137}c=\frac{h}{2\pi r_{S}}\times v_{S}\approx \frac{h}{2\pi r_{S}}\times c.\label{eq:ekinequi}
\end{equation}
Slater could not have achieved $r_{S}\approx 137a_{0}$ without assuming $\phi_{0}$ and was not able to figure out why we cannot let the $r_{S}$ increase to any desired size. By contrast, \cref{eq:ekinequi} answers them clearly without assuming $\phi_{0}$ anywhere. By comparing the left and right-hand sides of \cref{eq:ekinequi}, it is obvious that $r_{S}\approx 137a_{0}$. On the other hand, if an atomic size $r_{S}$ larger than $137a_{0}$ were possible, the equation for kinetic equilibrium, i.e., the left-hand side equation, would demand that the orbital velocity $v_{S}$ of such an atom should be larger than $c$, the speed of light. Such a situation is expressly forbidden in this physical world, in turn, this indicates that the radius of any atom in kinetic equilibrium never exceeds $137a_{0}$.
\vspace{1ex}

Some might notice that an extra negative charge was added to the core of the Slater's atom (\cref{fig3}d) by stealth. It was necessary to stabilize the Slater's atom electrically by all means. Yet where it comes from is another issue and has to be considered. The secret charge is discussed using the rest of this paper, together with the secret of $\alpha\approx \frac{1}{137}$ that sets the strict limit $137a_{0}$ on the size of an atom.

\section{Discussion}
\label{sec:disc}
\subsection{With respect to the fine structure constant $\alpha$}
\label{sec:fine}
As a dimensionless constant which does not seem to be directly related to any mathematical constant, the fine structure constant $\alpha=0.00729...~\approx \frac{1}{137}$ has long fascinated physicists~\cite{wikialpha}. There is a quote from Wolfgang Pauli: ``\emph{When I die my first question to the Devil will be: `What is the meaning of the fine structure constant?'}'' According to Richard Feynman, ``\emph{It's one of the greatest damn mysteries of physics: a magic number that comes to us with no understanding by humans.{\if0 You might say the `hand of God' wrote that number, and `we don't know how He pushed His pencil.'\fi}}''
\vspace{1ex}

The explicit quest for $\alpha$ was begun by Arnold Sommerfeld with his theory, which was hailed as a great progress, the final solution to the problem of the hydrogen atom and its spectrum~\cite{Kragh2003}. Let me begin by following his way, considering the ``\emph{atomic structure and spectral lines}''~\cite{SommerBook}.
\vspace{1ex}

The doubly superpositioned orbit gets the speed of light $c$ after the orbital expansion from the radius $a_{0}$ to $137a_{0}$ (\cref{fig3}c to d). This is the consequence of \cref{eq:ekinequi} assuming kinetic equilibrium, from de Broglie's wave view. Particle view, on the other hand, yields a different consequence. The $E_{kin}$ in particle view is given by \cref{eq:ekin}. Due to the expansion, the kinetic energy is lowered, hence, the system gains the amount of energy,
\begin{subequations}
\begin{align}
\Delta E_{kin}\equiv E_{kin}'-E_{kin}^{(S)} &\approx 2\times \frac{\hbar^{2}}{2ma_{0}^{2}}-2\times \frac{\hbar^{2}}{2m(137a_{0})^{2}}\\
&=2\times \frac{\hbar^{2}}{2ma_{0}^2}~\left(\frac{1}{1^{2}}-\frac{1}{137^{2}}\right)\\
&=2\times {\rm Ry}\left(\frac{1}{1^{2}}-\frac{1}{137^{2}}\right),\label{eq:deltaekin}
\end{align}
\end{subequations}
where the factor 2 reflects the current situation where two $\rm{h}^{+}$ particles take part in and the ${\rm Ry}$ is the Rydberg unit of energy, approximately 13.605 eV. \autoref{eq:deltaekin} suggests that the system after the expansion may save this amount of energy somewhere. However it is strictly forbidden. Kinetic \emph{equilibrium} demands that all the excess energy in \cref{eq:deltaekin} should be spent someway.
\vspace{1ex}

By the way, energy of a spectrum $h\nu$ emitted by a hydrogen atom is given by the Lyman series, one of the versions of the Rydberg formula,
\begin{equation}
h\nu = {\rm Ry}\left(\frac{1}{1^{2}}-\frac{1}{n^{2}}\right),\label{eq:lyman}
\end{equation}
where $n$ is a natural number greater than or equal to 2 (i.e., $n$ = 2, 3, 4, ...) and is now known as the principal quantum number. When an orbital charge falls from an initial energy level $n$ to the lowest energy level $n=1$, the atom emits radiation with the energy given by \cref{eq:lyman}. It is generally supposed that the number $n$ can take value infinity, and the special case ``$n=\infty$'' is called the Lyman limit. Contrary to the general supposition, however, the kinetic equilibrium in \cref{eq:ekinequi} sets the strict limit on $n$. If an orbit with a principal quantum number $n\geq 138$ were possible, the velocity of such an orbit would exceed the speed of light, which is expressly forbidden in this physical world where $c$ rules velocities of matter. Thus $n\geq 138$ is impossible, consequently, the maximum value for $n$ is 137.
\vspace{1ex}

Regarding $\alpha$, Paul Dirac in 1978 wrote: ``\emph{I doubt very much whether any really big progress will be made in understanding the fundamentals of physics until it is solved.}''~\cite{Kragh2003} So significant is the $\alpha$ that it seems highly impertinent for a mere human being to consider $\alpha$, even more to give it a physical meaning. Nonetheless, encouraged by Dirac's another teachings---``\emph{Scientific progress is measured in units of courage, not intelligence}'', I would like to give a physical meaning to $\alpha$---\emph{it defines the boundary between matter and the vacuum.} That is, a shell within $n=137$ is matter, and the outside of the shell is the vacuum. This conclusion seems very close to the Max Born's consideration: ``\emph{If $\alpha$ were bigger than it really is, we should not be able to distinguish matter from ether [the vacuum, nothingness] ... .}''~\cite{Miller2009} The outside of the shell is filled with \emph{ether}, where a charge cannot exist as matter. There go only electromagnetic waves, with the speed of light. $c$ does not indicate that $\infty$ is a good number for infinity. $c$ does indicate that $1/\alpha=137.035...$ is good enough for physical infinity.
\vspace{1ex}

This is the event taking place in superconducting transition. There once was an orbital charge, $\rm{h}^{+}$. When it encountered its partner, they were sent to the outer space beyond $n=137$ at the request of kinetic equilibrium. Then they have to convert themself to light. Because of their own ultimate speed, everything looks shrunk in size. Precisely written, everything looks as nothingness, and so does the core where they have just come from. They orbit around the core no matter what it is, no matter what fine structure it has. All they need is the information about the quantity of charges at the core.
\vspace{2ex}
\subsection{Pairing mechanism}
\label{sec:pair}
\begin{wrapfigure}{r}{0.43\textwidth}
\vspace{3ex}
\centering
\includegraphics[width=0.43\textwidth]{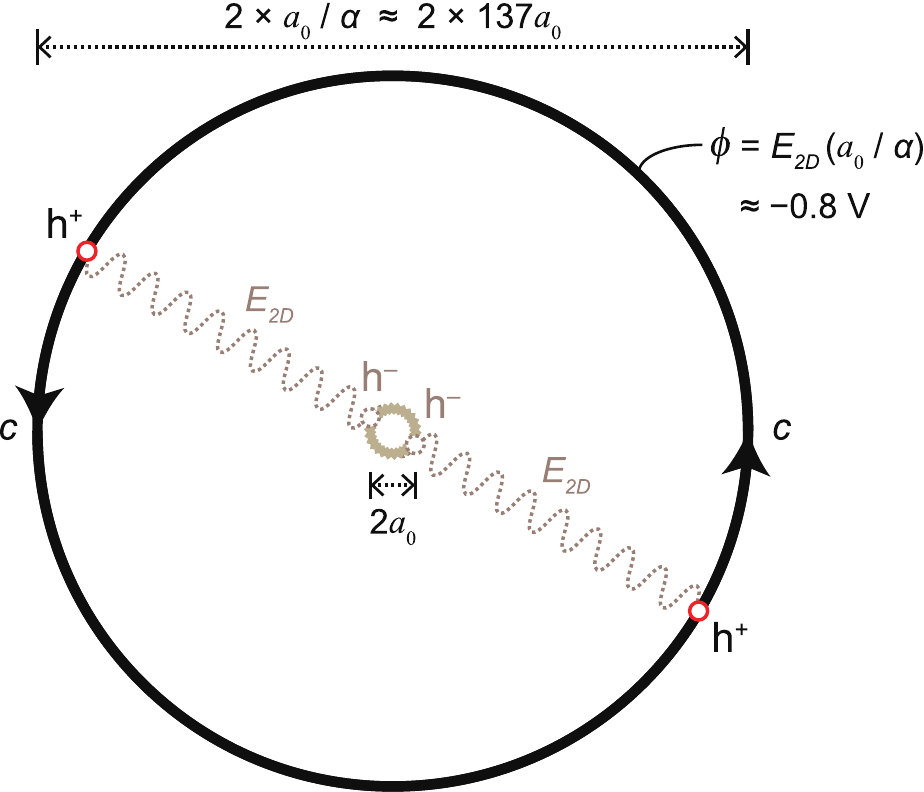}
\caption{\textbf{Superconducting atom and its fine structure.} An $\rm{h}^{+}$-pair having positive charges $+2q$ orbits around the core having negative charges $-2q$. The orbital radius and velocity is $a_{0}/\alpha\approx 137a_{0}$ and $c$, respectively. The $\rm{h}^{+}$-pair gains the `charge--light' duality with the help of its own antiparticle $\rm{h}^{-}$-pair being generated from the vacuum. The exchange takes place at the speed of light, i.e., at all times. The electric potential $\phi =E_{2D}(a_{0}/\alpha)\approx -0.8~\si{\volt}$ gives each orbital $\rm{h}^{+}$ the superconducting energy gap $\mathnormal{\Delta}$ of 0.8 eV; for details, see \cref{sec:gap}.
\label{fig4}}
\vspace{-3ex}
\end{wrapfigure}
Superconducting transition is such a pure physical event where a charge ($\rm{h}^{+}$) converts itself to light. The theory of quantum electrodynamics (QED) provides a coupling constant $\sqrt{\alpha}$ for the exchange and demands that its own antiparticle be generated from the vacuum. The antiparticle has a negative charge $-q$ in this case and goes back in time. The $\rm{h}^{+}$-pair propagating forward in time is what we usually recognize as a superconducting pair current. Due to its own `charge--light' duality, a pair of its own antiparticles---the $\rm{h}^{-}$-pair---is also necessary somewhere in the system.
\vspace{1ex}

Here we consider again both the extra negative charge added by stealth to the Slater's core (\cref{fig3}c to d) and the excess energy in \cref{eq:deltaekin} which has to be spent someway. Now, physical infinity is not $\infty$ but $1/\alpha$. Therefore the second term of \cref{eq:deltaekin} is precisely zero, and the excess energy can be rewritten exactly as
\begin{equation}
\begin{split}
\Delta E_{kin}=2{\rm Ry}=\frac{\hbar^{2}}{ma_{0}^2}=\frac{\hbar^{2}}{m}\times \frac{m^{2}q^{4}}{16\pi^{2} \epsilon_{0}^{2} \hbar^{4}}&=\frac{q^{4}}{4\epsilon_{0}^{2} h^{2} c^{2}}\times mc^{2}\\
&=\alpha^{2}\times mc^{2}.\label{eq:peace}
\end{split}
\end{equation}
For the fourth and sixth formula, the definition of $a_{0}$ in \cref{eq:a0} and $\alpha$ in \cref{eq:vbc} is used, respectively.
\vspace{1ex}

From Lyman's view, the $\Delta E_{kin}$ is equivalent to $h\nu$ emitted when a charge falls from $n=\infty$ to $n=1$. Now we already know $n=\infty$ is the vacuum. And the system still stores the excess energy $\Delta E_{kin}$. These indicate that the $\Delta E_{kin}$ can be used for inviting a charge from the vacuum to $n=1$. As written above, the vacuum is filled with electromagnetic waves, i.e., there is no space for a charge to exist. This is the essence of \cref{eq:peace}. That is, the equation gives an amount of energy that is necessary to invite an electromagnetic wave as a solid charge with mass, from the vacuum to the $n=1$ of the atom.
\vspace{1ex}

\noindent
\autoref{eq:peace} is not an inverted version of the Einstein's formula $E=mc^{2}$. The ``$E=mc^{2}$'' generates an enormous amount of \emph{energy} with the matter being completely destroyed whatever fine structure it has. By contrast, the ``$E=\alpha^{2}\times mc^{2}$'' generates a \emph{matter} having ``$m$'' in the atom with its fine structure being undestroyed, keeping peace in itself.
\vspace{1ex}

The superconducting atom is illustrated conclusively in \cref{fig4}. It has the orbital radius of $a_{0}/\alpha$ ($\gtrsim 137a_{0}$), meaning that the $\rm{h}^{+}$-pair is in the vacuum and hence is taking form of an electromagnetic wave with the speed of light. Rectilinear propagation, the tendency of an electromagnetic wave, has already gotten lost since the boundary between the matter and the vacuum is not homogeneous at all, any longer; there is no more reason to expect the same refractive index throughout.{\if0 This has just been defined by $\alpha$.\fi} At the core, there is the $\rm{h}^{-}$-pair. It stabilizes the superconducting atom electrodynamically. It goes back in time, hence, we cannot observe it usually{\if0; it is quite invisible\fi}. The $\rm{h}^{+}$-pair as well cannot observe it because of its own ultimate orbital velocity $c$, i.e., due to Lorentz contraction. And most importantly, all it needs is the information about the quantity of charges at the core. The excess energy $\alpha^{2}\times mc^{2}$ being stored before the kinetic expansion was used for inviting a single $\rm{h}^{-}$ from the vacuum to the orbital $n=1$. The $\rm{h}^{-}$-pair was thus formed at the orbital $n=1$, and the quantity of charges was thus equilibrated. Then there arises another mystery. What was the primitive negative charge at the atomic core before the kinetic expansion? In other words, what was the blue dot in \cref{fig3}c or b or a? It could be an anion Nb$^{-}$ or O$^{-}$ or NbO$^{-}$, or, it could be an antiparticle $\rm{h}^{-}$ already. This study began by setting two particles with opposite charges at a distance infinity (\cref{fig3}a). As written already, Newton's first law decrees that they each orbit the other whatever the other is, only if the other has the opposite charge. We now know the physical infinity is $1/\alpha$, indicating that if in space there are a particle and its antiparticle at a distance shorter than $1/\alpha$, they may form a metastable pair of orbits. The particle's electromagnetic shell may conceal the antiparticle's core. Also we cannot easily detect antiparticles which are always travelling backwards in time. Despite our limited ability to know and regardless of what is at the core, they will still orbit the other no matter how fine the structure they each have. The saga of $\alpha$ is still to be continued.
{}
\vspace{1ex}

To summarize, the superconducting atom as a whole is charge neutral. The $\rm{h}^{+}$-pair and its own anti-pair are coupled by $\alpha$ thanks to QED. As a consequence, they carry supercurrent with the speed of light $c$.

\subsection{London's canonical momentum}
\label{sec:phi}
\vspace{-1ex}
Bardeen demonstrated his respect for London's work by using his cut of the 1972 Nobel prize to fund the Fritz London memorial prize~\cite{LondonBro2011}. On the other hand, he used only a single page of the 30-pages long BCS paper~\cite{BCS1957} for the Meissner effect, seemingly disregarding London brothers' dedication and commitment to the Meissner effect~\cite{LondonBro1934}. As a matter of course, in the BCS paper published in 1957, there is no mention of the flux quantum ($\phi_{0}\equiv h/2q$) which was already predicted by London in 1948. In contrast to BCS, this study considers both $\phi_{0}$ and the Meissner effect.
\vspace{1ex}

London introduced the canonical momentum for superconducting electrons,
\begin{equation}
\vec{p}=m\vec{v}+q\vec{A},\label{eq:canop}
\end{equation}
where $m$ is the mass of the charge and $\vec{A}$ is the vector potential. Expecting superconducting state to have and retain zero net momentum $\left<\vec{p}\right>=0$, we then have the local average velocity $\left<\vec{v}_{s}\right>=-\frac{q}{m} \vec{A}$, which yields the current density $\vec{j}_{s}\equiv qn_{s}\left<\vec{v}_{s}\right>=-\frac{q^{2}n_{s}}{m} \vec{A}$, where $n_{s}$ is the number density of carriers. It is well known that the time derivative and the curl of $\vec{j}_{s}$ yield the famous London equations.
\vspace{1ex}

In normal state, a free point charge in a static field will move in a circle due to the Lorentz force. In this study, such a charge motion has been considered consistently as a wave having the kinetic momentum $mv=\frac{h}{2\pi r}$, where $r$ is an arbitrary radius of the circle.
\vspace{1ex}

When the point charge undergoes a superconducting transition, the radius $r$ expands to $a_{0}/\alpha$ ($\approx 137a_{0}$). By following London, $\left<\vec{p}\right>=0$. Given that the applied flux density is $\vec{B}=(0,0,\mu_{0}H)$, one possible solution for $\vec{\nabla}\times \vec{A}=\vec{B}$ is $\vec{A}=(0,\mu_{0}Hr,0)$. Now $r=a_{0}/\alpha$. We then have
\begin{equation}
0=\frac{h}{2\pi (a_{0}/\alpha)}+q\mu_{0}H\cdot a_{0}/\alpha,\label{eq:canopS}
\end{equation}
which yields
\begin{equation}
\mu_{0}H=-\frac{h}{2\pi q(a_{0}/\alpha)^{2}}=-\frac{h/2q}{\pi (a_{0}/\alpha)^{2}}.\label{eq:canopS2}
\end{equation}
This corresponds to \cref{eq:hc} that explains the experimental critical field. The London's canonical momentum thus yields the same conclusion given by Slater~\cite{Slater1937}, that a single diamagnetic $\phi_{0}$ is inherent in the superconducting atom.
\vspace{1ex}

By the way, it is noteworthy that what we have just considered above is neither a pair of charges nor a doubly superpositioned orbits. Nevertheless we have obtained the $\phi_{0}\equiv h/2q$, the indication of superconducting pairing. This must come from the assumption that $\left<\vec{p}\right>=0$. London interpreted it as the ``rigidity'' of the wavefunction of superconducting electrons. On the other hand \cref{eq:canopS2} indicates that it is associated with ``$2\pi$''. Without ``$2\pi$'', the pair ``$2q$'' cannot be generated. There once was a free charge moving in a circle. To make the circle ``rigid'', another charge was invited to the circle. Once they form a pair, the circle is expanded due to kinetic equilibrium. At the same time, the circle gets the ``rigidity'', i.e., $\left<\vec{p}\right>=0$, since their moving directions in the circle are always opposite to each other. The nature of a circle---$2\pi=6.28...$---thus contributes to the pairing. This is also true for the formation of the Bohr's atom as shown in the former part of this paper (\cref{fig3}a) where the 1D motions of two charges were converted into the pair of 2D orbital motions. Thus $2\pi$ is a significant physical number in addition to $\alpha$. Upon pairing, the dimension is always expanded. This must be the secret of the physical number $2\pi$.
\vspace{4ex}

Finally let me consider the superconducting transition from the perspective of the flux line. It once let a charge move in a circle using the Lorentz force and was thus penetrating the space surrounded by the circle. Upon superconducting transition, however, the stable flux line was entirely expelled from the space, furthermore from the space being expanding to the radius $a_{0}/\alpha$. During the event, the flux line cannot keep straight but is bending. That is, one source of the electromagnetic field is losing its fundamental tendency, rectilinear propagation. And the bending indicates that the magnetic flux line cannot observe anything in the circle any longer. According to Wolfgang Rindler, this is an ``\emph{event horizon}'' that he defined as ``\emph{a frontier between things observable and things unobservable.}''~\cite{Rindler1956} We have already studied that the bending is due to $\alpha$ that defines the boundary between matter and \emph{ether}. In terms of superconductivity, $\alpha$ sets the \emph{Meissner event horizon}.
\vspace{3ex}

Besides the flux line, external electromagnetic waves as well cannot invade the interior of the circle any more because the interior is the place only for matter. Whenever necessary to go inside, an electromagnetic wave has to convert itself to a charge using energy $E=\alpha^{2}\times mc^{2}$. If there is not enough energy, it cannot go inside but just bends, hence cannot obtain any information about the core of the superconducting atom, other than the information about the quantity of its charges. The bending motion of the flux line during the Meissner event induces another source of the electromagnetic field---the electric field---in accordance with Faraday's law of induction. Although the flux line was completely expelled from the interior of the superconducting atom, it alternatively left an electric field at the nearby atom. That is, the quantity of charges at the core is still observable and hence is still a matter of interest thanks to the Faraday's electric field, which is the main topic for the next last section.
\vspace{4ex}

The Higgs mechanism interprets the Meissner event as the generation of a mass $M$ for the magnetic field so that it bends. The $M$ is defined by $\lambda\equiv h/(Mc)$, the Compton wavelength of a quantum. In the case of superconducting transition, the $\lambda$ is identical with the penetration depth $\lambda_{\perp}=2(a_{0}/\alpha)\approx 2r_{S}$ which is proven by \cref{eq:lambda} in the former part of this paper. Then $M$ is
\begin{equation}
M=\frac{h}{\lambda_{\perp} c}=\frac{h\alpha}{2a_{0}c}=\frac{h}{2c}\times \frac{q^{2}}{2\epsilon_{0} h c}\times \frac{mq^{2}}{4\pi \epsilon_{0} \hbar^{2}}=\frac{q^{4}}{4\epsilon_{0}^{2} h^{2} c^{2}}\times \pi\times m=\pi\alpha^{2}\times m.\label{eq:higgs}
\end{equation}
For the fourth and sixth formula, the definitions of $a_{0}$ in \cref{eq:a0} and $\alpha$ in \cref{eq:vbc} are used. \autoref{eq:higgs} indicates that at the Meissner event horizon the bent flux line has the mass $M=\pi\alpha^{2}\times m$, which is approximately 5977 times smaller than the mass of a point charge. Whatever. The significance of \cref{eq:higgs} does not lie in the numerical result but in this,
\begin{equation}
\frac{M}{\pi\alpha^{2}}=m.\label{eq:gb}
\end{equation}
As discussed before, a good number for infinity in physics is not $\infty$ but $1/\alpha$. In turn, the physical infinitesimal is $\alpha$. Hence \cref{eq:gb} indicates that when a gauge boson at the event horizon is compressed into an infinitesimal circle with the area $\pi\alpha^{2}$, it may become a solid charge. What is most significant is that it is a \emph{2D circle} no matter how small it is. That is, the solid charge at the Meissner event horizon---the territory for superconductivity---is always a 2D particle.
\vspace{1ex}

There is a well-known reasonable supposition that 2D materials have advantage in getting superconductivity. In fact the 2D copper--oxygen plane is the bare essential of high-$T_{c}$ cuprate superconductors. \autoref{eq:gb} however does not pronounce 2D to be ``\emph{advantageous to superconductivity}'' but rather, it pronounces ``\emph{superconductivity is an event of 2D physics.}'' This is consistent with the experimental $H_{c}$ in \cref{eq:hc} and $I_{c}$ in \cref{eq:lambda} both of which are valid only in a 2D perspective.

\subsection{Faraday's electric field in the 2D world and Superconducting energy gap $\mathnormal{\Delta}$}
\label{sec:gap}
Hence the superconducting atom is a 2D atom. Therefore its electromagnetic shell has to be considered from a 2D perspective (see \cref{fig4}). Separated from kinetic energy and its equilibration, we here investigate its potential energy.
\vspace{1ex}

There are two $-q$'s at the core, and the electric field generated by them is given by Gauss's law in \cref{eq:gauss}. This time the shell is 2D, therefore, the surface integral is not $4\pi r^{2}$ but $2\pi r$. Then the 2D electric field is
\begin{equation}
E_{2D}(r)=-\frac{2q}{2\pi \epsilon_{0}} \frac{1}{r}=-\frac{q}{\pi \epsilon_{0}} \frac{1}{r}.\label{eq:e2d}
\end{equation}
Note the unit of $E_{2D}(r)$ is already $\frac{[\si{\coulomb}]}{[\si{F/m}][\si{m}]}=[\si{\volt}]$. What we are considering now is not a 2D world which is merely uniform in the $z$-direction but the perfect 2D world where we cannot assume the charge per unit length in the $z$-direction. If we assume it, the unit of \cref{eq:e2d} would be $[\si{V/m}]$, then, the formula $\vec{E}(r)\equiv -\rm{grad}~\it{\phi(r)}$ would give a logarithmic electric potential $\phi(r)$ in the unit $[\si{V}]$. However, as proven in the previous section, any charge within the superconducting atomic shell is a perfectly 2D particle. Hence such a charge per $z$-directional unit length is not a valid assumption at all. In other words the unit of $q$ in \cref{eq:e2d} should not be taken as $[\si{C/m}]$ but as $[\si{C}]$. In turn, $[\si{V}]$ as the unit of $E_{2D}(r)$ is correct. The electric field in the 2D world might be regarded as the wheat field being completely flattened out against the ground. There is no ``wind'' blowing across the ``field''. The field lying down on the ground itself shows how strong the wind is.
\vspace{1ex}

Additionally, the electric field does not exist in the interior of the 2D superconducting shell but does exist only within a thin layer adjacent to the periphery, i.e., only within the Meissner horizon. If we could draw a Faraday's line of electrostatic force connecting $-q$ at the core to $+q$ at the periphery, $\phi(r)$ would be easily obtained according to $\vec{E}(r)\equiv -\rm{grad}~\it{\phi(r)}$. Obviously this is impossible. We cannot explicitly draw a Faraday's line in the superconducting shell. But still, those $-q$ and $+q$ are invisibly connected to each other by $\sqrt{\alpha}$ as demanded by QED. This is due to the fact that those particles are antiparticles of the other. No matter how far they are being coupled. Although the electric field is no longer taking the form of a vector, it seems possible to define the electric potential at the two separate points, $r=a_{0}$ and $a_{0}/\alpha$, thanks to $\alpha$. Whether this situation is still under Faraday's principle of locality or under Newton's action at a distance is just as staring into the abyss.
\vspace{1ex}

Concluding the above arguments, the electrostatic potential of the 2D superconducting shell is given by the Dirac delta function as
\begin{equation}
\phi=\int E_{2D}(r)\delta(r-a_{0}/\alpha)dr=E_{2D}(a_{0}/\alpha)=-\frac{q}{\pi \epsilon_{0}} \frac{1}{a_{0}/\alpha}\approx -0.794~\si{\volt}.\label{eq:V}
\end{equation}
This is a universal quantity and shall give each $\rm{h}^{+}$ at the shell the superconducting energy gap $\mathnormal{\Delta}$ of approximately 0.8 eV. This will be confirmed in the next experimental paper studying the macroscopic superconducting properties of this sample as the 2D JJ square array (\cref{fig1}b).
\vspace{3ex}

By the way, potential energy of the 2D atom just before kinetic expansion has not yet been calculated so far (see \cref{fig3}c). There was a single negative charge at the core and there were two positive charges at the unstable $[\rm{h}^{+}\!\times\rm{h}^{+}]$-orbit. Any potential energy possesses significant information about the physical system, particularly about the steady state of the system. For example, a potential energy expressed as $-G\frac{M_{1}M_{2}}{r}$ where $G$ is the gravitational constant indicates that the attractive gravitational force governs the system and that there are two bodies in the system. Now according to \cref{eq:e2d} the electric potential at the unstable $[\rm{h}^{+}\!\times\rm{h}^{+}]$-orbit turned out to be $-\frac{q}{2\pi \epsilon_{0}}\frac{1}{a_{0}}$. Therefore its potential energy just before kinetic expansion was
\begin{equation}
E_{pot}'=2q\times -\frac{q}{2\pi \epsilon_{0}}\frac{1}{a_{0}}=-\frac{q^{2}}{\pi \epsilon_{0}}\frac{1}{a_{0}}=-\frac{q^{2}}{\pi \epsilon_{0}}\frac{mq^{2}}{4\pi \epsilon_{0} \hbar^{2}}=-\frac{q^{4}}{\epsilon_{0}^{2} h^{2} c^{2}}\times mc^{2}=-4\alpha^{2}\times mc^{2}=-4\Delta E_{kin}.\label{eq:epot'}
\end{equation}
For the fourth, sixth, and seventh formula, the definition of $a_{0}$ in \cref{eq:a0}, $\alpha$ in \cref{eq:vbc}, $\Delta E_{kin}$ in \cref{eq:peace} is used, respectively. As considered in \autoref{sec:pair}, the $\Delta E_{kin}$ corresponds to the energy necessary to invite an electromagnetic wave as a solid charge from the vacuum to the inner atom. Hence \cref{eq:epot'} indicates that the system before kinetic expansion already possesses enough potential for realizing a four-body system inside the superconducting atom. That is, this study claims that superconductivity is not by a charge pair but by a charge quartet. The quartet consists of a particle $\rm{h}^{+}$-pair and an antiparticle $\rm{h}^{-}$-pair, as already illustrated in \cref{fig4}. The former moves forward in time, and the latter moves backward in time. They each have opposite charges to each other. By taking into account the definition of current $I\equiv dQ/dt$, both pairs contribute to the amount of supercurrent, thus, it turns out that \cref{eq:ic} is conclusively correct.

\section{Conclusions}
\label{sec:conc}
If there are two particles with opposite charges in space, they \emph{absolutely} orbit around the other---this is what Newton's first law is indicating. If the orbit is superpositioned and gets two times larger kinetic momentum, it \emph{absolutely} gets unstable and is \emph{absolutely} going to expand---the same as celestial mechanics. And the expanded orbital radius \emph{never} exceeds $a_{0}/\alpha\approx 137a_{0}$ because the velocity of any matter in this physical world \emph{never} exceeds the speed of light $c$.

\vspace{1ex}
Those three deterministic physics---orbiting, orbital expansion, and the orbital ultimate velocity---are all that this study needed to explore superconducting transition. Cooper~\cite{Cooper1956} constructed a wave function for a single pair of electrons excited above the Fermi surface and found that for a negative interaction a bound state is formed~\cite{BCS1957Letter}; yet it does not generate the London and Slater's superconducting atom. This study on the other hand did not require such a wave function for any pair, whether attraction or repulsion, from the start to the end; nevertheless, the superconducting atom is successfully generated.

\vspace{1ex}
This study has paid attention not only to the London and Slater's superconducting atom but also to the $\alpha$ which Sommerfeld, Born, Pauli, Dirac and Feynman have paid much attention to. And this study has perceived that the $\alpha$ sets strict boundaries between atoms and the vacuum in an inanimate matter. On the other hand, BCS~\cite{BCS1957} followed Klein and Lindhard who fixed their attention on the behaviour of what they called ``\emph{an infinite homogeneous medium}''~\cite{Klein1945}. Infinite homogeneity however cannot define such definite boundaries or $\alpha$. Hence, it cannot create even a single atom in its interior, and consequently nowhere does it form the London and Slater's atom. The superconducting atom and the BCS ground state are thus mutually exclusive, and so are this study and the BCS theory.

\vspace{1ex}
Thanks to Hirsch's kinetic expansion together with de Broglie's wave view, this study has succeeded in generating a single flux quantum in the superconducting atom through the London's canonical momentum. Additionally, by considering the Meissner effect through the Higgs mechanism, this study has revealed that 2D materials are not just advantageous to superconductivity but rather that \emph{superconductivity is an event of 2D physics}. Now the answer for the famous question why $T_{c}$ is high for cuprates and is low for metals becomes clear---this is because the former has inherent \emph{2D} copper--oxygen planes in its own structure but the latter---the aggregation of \emph{3D} grains---does not have such a \emph{2D} structure. It is surprising that such a simple explanation drawn from the Higgs mechanism can be employed to answer the long-standing question.

\vspace{3ex}
It turns out that superconducting transition is a pure physical event where an $\rm{h}^{+}$-pair converts itself to light. Quantum electrodynamics demands its antiparticle $\rm{h}^{-}$-pair to be generated from the vacuum. Among the quartet, the first half moving forward in time is what we usually recognize as a superconducting pair. Its well-recognized bosonic behaviour is a matter of course since the charge pair is taking the form of an electromagnetic wave---in particle view, a photon---at the same time. That is, the pairing is not related to pho{\bf n}ons but rather to pho{\bf t}ons. But still, external pairing glue is unnecessary. The $\rm{h}^{+}$-pair is mediated by itself as a gauge boson, thanks to the $\rm{h}^{-}$-pair of its own.

\vspace{1ex}
Hence the phonon-mediated ``\emph{Cooper pair}'' has nothing to do with the superconducting atom and is unrelated to the physics of this study. Alternatively ``\emph{Hirsch pair}'' will pronounce all appropriately~\cite{Hirsch2022APL,Hirsch2020APS}. This study has conclusively predicted that a universal quantity for the superconducting energy gap $\mathnormal{\Delta}$ is to be 0.8 eV. One of the most significant BCS formulae $\mathnormal{\Delta}=1.76k_{B}T_{c}$ then predicts $T_{c}\approx$ 5270 K exceeding the boiling point (BP) of the sample material used for this study, Nb having the BP of 5020 K. Despite the BCS theory's popularity, this study has demonstrated that superconductivity does occur without the BCS theory. If the future study proves $\mathnormal{\Delta}\approx$ 0.8 eV experimentally, the $\mathnormal{\Delta}$ itself will validate just that. Dawn of a new era begins with a quest for the fine structure at the core of a superconducting atom. There must be new physics still waiting to be found.
\vspace{5ex}

\noindent
{\small
{\bf Conflict of Interest:}
None declared.
\\
{\bf External Funding:}
None received.
\\
{\bf Data Availability:}
All experimental raw data are deposited at Zenodo~\cite{Zenodo2}.
}
\urlstyle{same}

\vspace{3ex}
\appendix
\section{Staring into the atomic core using $\pi$ and $\alpha$}
\label{sec:appA}
The Higgs mechanism yielded the mass of a bent magnetic flux line at the Meissner event horizon as $M=\pi\alpha^{2}\times m$. This concept represented by \cref{eq:gb} is applicable not only to the mass but also to a length scale because both $\pi$ and $\alpha$ are dimensionless. In this section, we stare at the Bohr's core. As the superconducting atom appears with the orbital expansion, it may be plausible to assume that the usual Bohr's atom having the orbital radius $a_{0}$ is also a consequence of some sort of expansion. If it was a 2D expansion, the origin should have the radius,
\begin{equation}
r_{O}=\frac{a_{0}}{\pi(1/\alpha)^{2}}\approx 0.897\times 10^{-15}~\si{m},\label{eq:ro}
\end{equation}
where $\pi(1/\alpha)^{2}$ is the area of an infinite 2D circle in terms of physics. That is, we have just assumed that an infinite 2D expansion of something at the origin with the radius $r_{O}$ resulted in the formation of the Bohr's atom having the radius $a_{0}$. This time the numerical result is somewhat important because we already know the radius of a proton is approximately~\cite{sci2019,nat2019} 0.83$\sim$0.88$\times 10^{-15}~\si{m}$. The rough coincidence, however, does not necessarily indicate that the original particle was a proton. $\pi$ and $\alpha$ merely indicate a possibility that the Bohr's atom was once upon a time a certain particle having the radius $r_{O}$. They do not indicate what it was. They just indicate it had the radius $r_{O}$ no matter which elementary particle it was.{\if0 This is determined by $\pi$ and $\alpha$.\fi}
\vspace{2ex}

On the other hand, we know the Bohr orbital velocity is not at the speed of light. Therefore, the assumed 2D expansion must be different from the kinetic expansion that yields the ultimate orbital velocity $c$. Also, we know the Bohr's atom is in the form of 3D. Nevertheless, \cref{eq:ro} indicates that the Bohr's atom is still in the form of 2D. In order to overcome this discrepancy, let me assume that once upon a time the ancient particle upgraded its own dimension using $\pi$. As already discussed in \cref{sec:phi}, the physical number $\pi$ is responsible for the dimension expansion. This time the dimension expansion is not associated with pairing, therefore, we use $\pi$ instead of $2\pi$. That is, we assume that at some point the ancient particle having the radius $r_{O}$ evolved into a higher-dimension particle with the orbital radius $\pi r_{O}$.
\vspace{4ex}

By the way, there is another significant radius in the range between $\pi r_{O}$ and $a_{0}$. According to \cref{eq:ekinparticle}, the orbital velocity of an arbitrary orbit with the radius $r$ is $v=\frac{\hbar}{mr}$. Hence, when $r=\frac{\hbar}{mc}$, the orbital velocity achieves the speed of light $c$. The ratio of such an $r$ to $a_{0}$ is
\begin{equation}
\frac{r}{a_{0}}=\frac{\hbar}{mc}\times \frac{mq^{2}}{4\pi \epsilon_{0} \hbar^{2}}=\frac{q^{2}}{4\pi \epsilon_{0}\hbar c}=\alpha.\label{eq:ao}
\end{equation}
For the second and fourth formula, the definition of $a_{0}$ in \cref{eq:a0} and $\alpha$ in \cref{eq:vbc} is used, respectively. The superconducting atom with the orbital velocity $c$ appears with the kinetic equilibrium expanding its orbital radius from $a_{0}$ to $a_{0}/\alpha$. \autoref{eq:ao}, on the other hand, indicates that the ultimate orbital velocity is also achievable when the orbital radius shrinks from $a_{0}$ to $a_{0}\alpha$. For such a small orbit having the radius $a_{0}\alpha$, the elementary charge $q$, its mass $m$, and $\alpha$ may not be good numbers any longer to describe its physics. In fact, at smaller distances, $\alpha$ is becoming larger. For example, at the mass scale of about 91 GeV/$c^{2}$ of the {\fontencoding{T1}\fontfamily{put}\selectfont Z} boson, an effective $\tilde{\alpha}$ is in the $\frac{1}{127}\sim\frac{1}{128}$ value range~\cite{Langacker2002,Frit2002arxiv} instead of $\frac{1}{137}$. Because of those mass-scale dependent variables, it is difficult to calculate accurately the special radius which yields the orbital velocity $c$. Regardless of what the accurate radius may be, what matters most is the fact that the special orbit can exist inside the Bohr's atom.
\vspace{2ex}

What realizes the ultimate orbital velocity $c$ is the kinetic equilibrium as already studied in superconducting transition; a doubly superpositioned orbit absolutely gets unstable in itself, then absolutely expands itself, and ultimately achieves the speed of light $c$. During the event, the excess kinetic energy expressed as $\Delta E_{kin}=\alpha^{2}\times mc^{2}$ according to \cref{eq:peace} has to be spent someway, and it has been done so by inviting an $\rm{h}^{-}$ charge from the vacuum to the core of the atom.
\vspace{2ex}

By taking into account this kinetic equilibrium principle, the possible existence of the special orbit having $c$ inside the Bohr's atom itself indicates that once upon a time the ancient orbit having the radius $\pi r_{O}$ was doubly superpositioned by some means---then the orbital radius expanded to approximately $\pi r_{O}\times 1/\tilde{\alpha}$ ($\approx a_{0}\tilde{\alpha}$), and the ultimate orbital velocity $c$ has thus been achieved. As with the superconducting transition, the excess kinetic energy had to be spent someway, and a certain charged particle had to be invited from the vacuum to the core. We of course know the charged particle at the core of the Bohr's atom is a proton. That is, the assumed charge invited from the vacuum during the kinetic expansion should be a proton. Then, the amount of the excess energy spent during the event can be expressed as
\begin{equation}
\Delta E_{kin}=\tilde{\alpha}^{2}\times m_{p}c^{2},\label{eq:ekinmp}
\end{equation}
where $m_{p}$ is the mass of a proton, approximately $1.673\times 10^{-27}~\si{\kilogram}$. Here we have used $\tilde{\alpha}\approx \frac{1}{127}\sim\frac{1}{128}$ instead of $\alpha\approx\frac{1}{137}$ because the orbit of interest is at the nearby nucleus. Hence the radius of the ancient particle in \cref{eq:ro} should also be rewritten as
\begin{equation}
\tilde{r_{O}}=\frac{a_{0}}{\pi(1/\tilde{\alpha})^{2}}.\label{eq:rotilde}
\end{equation}
\vspace{4ex}

On the other hand, the excess kinetic energy can be expressed in a different way as $\Delta E_{kin}=\frac{\hbar^{2}}{mr^2}$ according to \cref{eq:peace}.
Given that before kinetic expansion there once was a single particle with the mass $m_{O}$ in the ancient orbit having the radius $\pi \tilde{r_{O}}$, the excess energy accumulated due to double superpositioning was
\begin{equation}
\Delta E_{kin}=\frac{\hbar^{2}}{m_{O}(\pi \tilde{r_{O}})^{2}}.\label{eq:ekinmo}
\end{equation}
Again, this excess energy was spent inviting a proton. Hence, by equating \cref{eq:ekinmp} and \cref{eq:ekinmo}, we obtain
\begin{equation}
m_{O}=\frac{\hbar^{2}}{(\pi \tilde{r_{O}})^{2}}\times \frac{1}{\tilde{\alpha}^{2}\times m_{p}c^{2}}=\frac{\hbar^{2}}{a_{0}^{2}\tilde{\alpha}^{6}}\times \frac{1}{m_{p}}\times \frac{1}{c^{2}}\approx 62.18\sim65.18~\si{GeV}/c^{2},\label{eq:mo}
\end{equation}
where $\frac{1}{127}\sim\frac{1}{128}$ is used for $\tilde{\alpha}$, respectively.
\vspace{4ex}

After kinetic expansion, a special orbit has been formed at the nearby core. Its orbital radius is roughly $127\sim128$ times larger than the core and is roughly $127\sim137$ times smaller than the Bohr orbital. In particle view, the special orbit consists of a pair of particles, and the pair is twining around the core proton with the speed of light $c$. Each particle has the mass $m_{O}$, that is, the total mass of the special orbit is $2m_{O}\approx 124\sim130~\si{GeV}/c^{2}$, which corresponds to the experimentally measured mass of a Higgs boson~\cite{Ellis2012,ATLAS,CMS,PDG2022}. It may be a mere coincidence, yet such a twining motion of a particle pair having the speed of light always seems to generate the mass as does the Hirsch pair at the Meissner event horizon.
\vspace{5ex}

\begin{center}
{\large\bf S}{\small\bf UMMARY}
\end{center}
From an atomic origin far, far away{\if0---physically $\infty$\fi}, there is a Higgs event horizon{\if0 orbiting around the nucleus with the speed of light\fi}. From the Higgs event horizon far, far away{\if0---physically $\infty$\fi}, there is a Bohr orbital. And from the Bohr orbital far, far away{\if0---physically $\infty$\fi}, there is a Meissner event horizon{\if0 orbiting around the Bohr orbital as the core of the superconducting atom with the orbital velocity $c$\fi}. Both horizons orbit around the core with the velocity $c$ no matter what fine structure the core has.
{}

\end{document}